\documentclass[10pt,epsfig]{article}

\usepackage{multicol}
\usepackage{graphicx}
\usepackage{booktabs}
\usepackage{amssymb,bm,mathrsfs,bbm,amscd}
\usepackage[tbtags]{amsmath}
\usepackage{lastpage}
\usepackage{lscape}
\usepackage[rotateright]{rotating}

\begin{document}
\title{Projectile fragment emission in the fragmentation of $^{56}$Fe on Al, C and CH$_{2}$ targets}

\author{Luo-Huan Wang$^{1}$, Liang-Di Huo$^{1}$, Jia-Huan Zhu$^{1}$, Hui-Ling Li$^{1}$, Jun-Sheng Li$^{1}$\\
 S. Kodaira$^{2}$, N. Yasuda$^{3}$, Dong-Hai Zhang$^{1}$\thanks{Corresponding author. Tel: +863572051347; fax: +863572051347. E-mail address:zhangdh@sxnu.edu.cn}\\
$^{1}$Institute of Modern Physics, Shanxi Normal University, Linfen 041004, China\\
$^{2}$Radiation Measurement Research Section, National Institute of Radiological Science\\
Chiba 263-8555, Japan\\
$^{3}$Research Institute of Nuclear Engineering, University of Fukui, Fukui 914-0055, Japan}

\date{}
\maketitle

\begin{center}
\begin{minipage}{140mm}
\vskip 0.4in
\begin{center}{\bf Abstract}\end{center}
{The emission angle distribution of projectile fragments (PFs) and the temperature of PFs emission source for fragmentation of $^{56}$Fe on polyethylene, carbon and aluminum targets at the highest energy of 496 A MeV are investigated using CR-39 plastic nuclear track detector. It is found that the averaged emission angle of PFs increases with the decrease of PF charge for the same target, and no obvious dependence of angular distribution on the mass of target nucleus is found for the same PF. The cumulated squared transverse momentum distribution of PF can be well represented by a single Rayleigh distribution, the temperature of PFs emission source is extracted from the distribution, which is about $1.0\sim8.0$ MeV and do not depend on the mass of target for PF with charge of $9\leq{Z}\leq25$.}\\

\end{minipage}
\end{center}

\vskip 0.4in
\baselineskip 0.2in
\section*{1. INTRODUCTION}

Heavy ion collisions at intermediate and high energy, being an idea tool for producing hot nuclear matter at various densities in the laboratory, is an important source of information on the properties of nuclear matter under extreme conditions[1,2]. Experimental studies in this field extend from the Fermi energy regime to relativistic energies. Such study can help to understand not only the fundamental nuclear physics processes involved in nuclear fragmentation but also the thermodynamic evolution in the collisions. Multi-fragmentation is a universal phenomenon occurring when a large amount of energy is deposited in a nucleus. At low excitation energies, the produced nuclear system can be treated as a compound nucleus[3] which decays via evaporation of light particles or fission. However, at high excitation energy, a considerable amount of excitation energy and a slight momentum transfer are induced, possibly accompanied by compression during the initial dynamical stage of the interaction[4,5], the system will expand to subsaturation densities, thereby becoming unstable, and will break up into many fragments.

One of the major interests in the study of intermediate and high energy heavy ion collisions is the understanding of multi-fragmentation phenomenon and its connection with liquid-gas phase transition[2]. For this it has to be assumed that in a heavy ion collision at some stage a part of the system is both in thermodynamical equilibrium and instable. Such a configuration is often termed a freeze-out configuration. The multi-fragmentation process would reflect the parameters of this source, {\em i.e.} its temperature, density, and perhaps collective flow pattern. Among these parameters, temperature is an important thermodynamic quantity in the nuclear equation of state which is of broad interest in heavy ion collisions, since the nuclear system experiences an evolution from a very high temperature to a very low one to form the final fragments. Experimentally it is very difficult to determine the temperature of hot nuclear matter in a dynamical process. There are several nuclear thermometers have been proposed to extract the temperature. These include the slope of energy spectra[6-8], momentum fluctuations[9], double isotope yield ratios[10-13], and excited state distributions[14] among others. However they may not be generally applicable in all circumstances and even for a given system the extracted temperature value from these thermometers may be quite different from each other[15]. The typical temperature extracted from isotope ratios or level population ratios are $5\sim8$ MeV[11-13,16], which is in agreement with that used in statistical models to reproduce the experimental mass yield curves. The typical slope temperature extracted from kinetic energy spectra of PFs is about 17 MeV[8].

In this paper, the experimental results for the emission angle distribution of PFs and the temperature of PFs emission source obtained for fragmentation of  $^{56}$Fe on Al, C and CH$_{2}$ targets at highest energy of 496 A MeV are investigated using CR-39 plastic nuclear track detector.

\section*{2. EXPERIMENTAL DETAILS}

Stacks of Al, C and CH$_{2}$ targets sandwiched with CR-39 plastic nuclear track detectors (HARZLAS TD-1, Fukuvi, Japan) were perpendicularly exposed to a $^{56}$Fe beam of initial energy of 500 A MeV in the biology port of the HIMAC (the Heavy Ion Medical Accelerator in Chiba)facility at the Japanese National Institute of Radiological Sciences(NIRS). The beam fluence is about 2000 ions/cm$^{2}$. The configuration of sandwiched target is shown in Fig.1. A CR-39 sheet, with dimension of 5cm$\times$5cm and thickness of 0.8 mm, were placed before and after each targets. The thickness of aluminum, carbon and polyethylene targets is 2, 5 and 10 mm, respectively. There are two targets in each stacks, one CR-39 sheet placed before and after each target. The beam energy on each target upper surface were calculated using SRIM-2008 simulation code. The beam energy on each first target (target 1) upper surface is 496 A MeV, on second Al target (target 2) upper surface is 477 A MeV, on second C target (target 2) upper surface is 468 A MeV,  and on second CH$_{2}$ target (target 2) upper surface is 462 A MeV. The accuracy of beam energy on each target is calculated based on the uncertainty of length of beam line and the uncertainty of CR-39 sheet, which is less than 1 MeV/nucleon.

\begin{figure}[htbp]
\begin{center}
\includegraphics[width=0.60\linewidth]{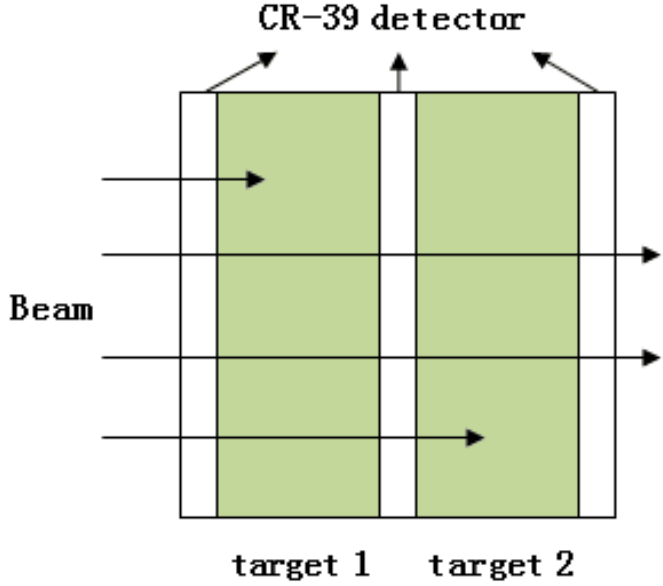}
\caption{(Color online) Sketch of the target-detector configuration.}
\end{center}
\end{figure}

After exposure, the CR-39 detectors were etched in 7 N NaOH aqueous solution at temperature of 70$^{\circ}$ for 30 hours. Tracks from beam ions and their fragments manifest in the CR-39 sheet as etch-pit cones on both sides of sheets. The images of ions tracks were scanned and analyzed automatically by HSP-1000 microscope system and the PitFit track measurement software, then checked manually one by one. When the track is mis-identified in the automatical PitFit track measurement process(such as target fragment track, disguised track, and overlapped track), the corresponding track is refitted in the manual checking process. The manual checking process is really a time consumed process, but it keeps the detection efficiency at 100$\%$. The PitFit software provide some geometric information of each track, such as the track coordinates, major and minor axes and area of etched track spot on CR-39 sheet surfaces. About 1.5$\times10^{4}$ $^{56}$Fe ion tracks were traced from the first CR-39 detector upper surface in the stack. The spots on the front surface (with respect to the beam direction) were directly scanned firstly, then the sheet were turned under the middle line of the sheet and the spots on the back surface were scanned.

Beam $^{56}$Fe trajectories and ones of secondary projectile fragments were reconstructed in the whole stack using the track tracing method[17]. First, the track position in CR-39 sheet surface is corrected by parallel and rotational coordinate transition. Second, the difference between the track position of corresponding tracks on both side of the CR-39 sheets and on the surfaces neighboring targets are minimized by a track matching routine.

The coordinate of the track before the target (or upper surface of CR-39 sheet) is $(x, y)$ and of the matching track after target (or back surface of CR-39 sheet) is $(x', y')$. Following the translation relation, the coordinate of matching track can be calculated as:

\begin{eqnarray}
x'_{th}&=&ax+by+c \nonumber \\
y'_{th}&=&a'x+b'y+c'
\end{eqnarray}
where $a, b, c, a', b'$, and $c'$ are fitting parameters, which can be fitted using the least square methods by choosing the coordinate of at least 100 $^{56}$Fe beam track. The calculated coordinate $(x'_{th}, y'_{th})$ of matching track is different from the measured $(x', y')$, the difference $dx=x'_{th}-x'$, $dy=y'_{th}-y'$ can be used to determine the matching track.

\begin{figure}[htbp]
\begin{center}
\includegraphics[width=0.60\linewidth]{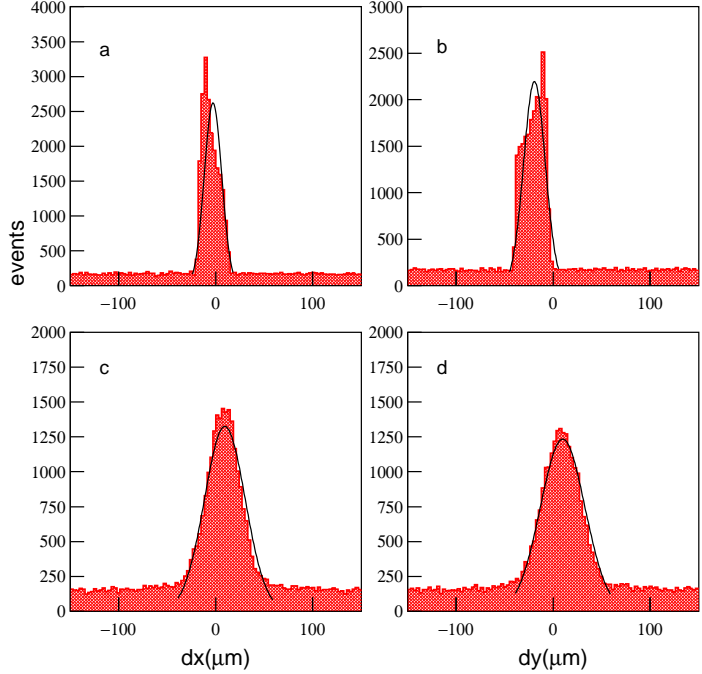}
\caption{(Color online) The track coordinate difference dx and dy distribution for 496 A MeV $^{56}$Fe fragmentation on CH$_{2}$ target, a and b the difference between the front and back surface on CR-39 sheet, c and d the difference before and after target.}
\end{center}
\end{figure}

Fig.2 shows the coordinate differences dx and dy distribution for 496 A MeV $^{56}$Fe fragmentation on CH$_{2}$ target, a and b are the track coordinate difference in the upper surface and down surface on the sheet, c and d are the track coordinate difference before and after the target, the smooth line in each figures is the Gaussian fits of the distribution. The difference are calculated for all combinations of positions for extracted tracks, only the matched combination ought to make a peak which appears in the figures, the dx and dy values of other combinations should be randomly distributed. The deviations ($\sigma(dx)$ and $\sigma(dy)$) give the position accuracies of track which are estimated to be 3.0$\sim$12.0 $\mu$m between the upper and down surface of a CR-39 sheet and 6.0$\sim$30.0 $\mu$m before and after targets which depend on the thickness of CR-39 sheet, type of target and thickness of target. The accuracy suffers from Coulomb scattering becomes greater on the downstream detectors because Coulomb scattering effect increases with the increase of the energy loss. However, since the matching track is searched within four times of the deviations in our investigation, the Coulomb scattering effect is negligible.

There are several possibilities when matching tracks in the region of ($x'_{th}\pm4\sigma(dx)$, $y'_{th}\pm4\sigma(dy)$) or in the region of projectile fragmentation angle ($\theta_{fr}=p_{f}/p_{beam}$, where $p_{f}$ is the Fermi momentum of nucleon in nuclei which is about 200 MeV/c, $p_{beam}$ is the momentum of beam $^{56}$Fe) on the detector surfaces which are adjacent to the target. In present experiment we choose the region ($x'_{th}\pm150\mu{m}$, $y'_{th}\pm150\mu{m}$) as a candidate track searching region, which is larger than the region of projectile fragmentation angle.

(1) There is a candidate track in the region of ($x'_{th}\pm150\mu{m}$, $y'_{th}\pm150\mu{m}$) and the area of the matched track is in the region of beam $^{56}$Fe track area, which is considered as a no fragmentation event.

(2) There is a candidate track in the region of ($x'_{th}\pm150\mu{m}$, $y'_{th}\pm150\mu{m}$) and the area of the matched track is lesser than the region of beam $^{56}$Fe track area, which is considered as a fragmentation event.

(3) There are two or three candidate tracks in the region of ($x'_{th}\pm150\mu{m}$, $y'_{th}\pm150\mu{m}$) and the sum of the matched track charge is equal or lesser than the charge of beam $^{56}$Fe, which is also considered as a  fragmentation event.

(4) There is a candidate track in the region of ($x'_{th}\pm150\mu{m}$, $y'_{th}\pm150\mu{m}$) and the area of the matched track is greater than the region of beam $^{56}$Fe track area, which is considered as a charge-pickup reaction event.

(5) There is no candidate track in the region of ($x'_{th}\pm150\mu{m}$, $y'_{th}\pm150\mu{m}$), it is considered as completely fragmentation event, but the charge of fragments are below the threshold of CR-39 detector.

The charge of produced projectile fragment is determined from the etched track area distribution. Fig.3 shows the etched track area distribution of $^{56}$Fe and their fragments for 462 A MeV $^{56}$Fe fragmentation on CH$_{2}$ target (second target). The beam $^{56}$Fe and their fragments with charge up to $Z=20$ is clearly shown as peaks. For other fragments with charge less than $Z=20$, it can not be identified directly from the figure. Using seven Gaussian superposition fitting we can get the mean etched track area and its deviation of $^{56}$Fe and their fragments with charge greater than $Z=19$. Fig.4 shows the dependence of etched track area on the charge from Gaussian simulated results in Fig.3 for $^{56}$Fe and their fragments with charge greater than $Z=19$. The etched track area increases linearly with increase of the charge ($Z\geq20$) of fragments up to beam $^{56}$Fe. Using linear fitting we can get $Area=33.39+279.36Z$ with $\chi^{2}_{min}=0.01$. Based on this dependence the charge of other fragments can be determined.

\begin{figure}[htbp]
\begin{center}
\includegraphics[width=0.60\linewidth]{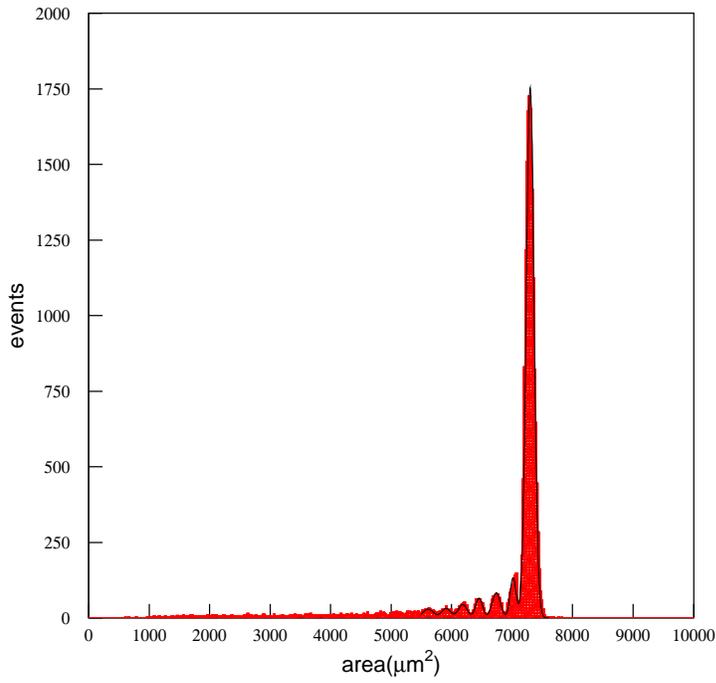}
\caption{(Color online) The etched track area distribution of $^{56}$Fe and their fragments for 462 A MeV $^{56}$Fe fragmentation on CH$_{2}$ target, the smooth line is the Gaussian fit.}
\end{center}
\end{figure}

\begin{figure}[htbp]
\begin{center}
\includegraphics[width=0.60\linewidth]{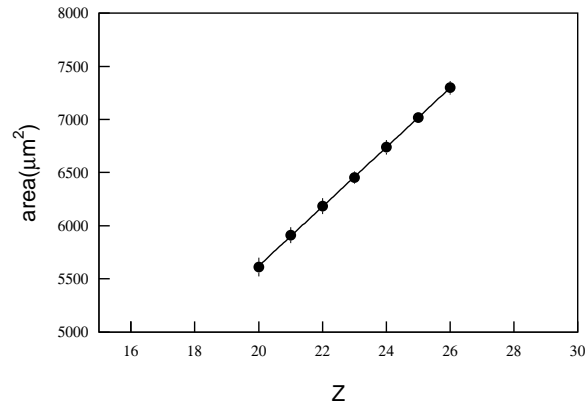}
\caption{The dependence of etched track area on the charge from Gaussian simulated results in Fig.3 for $^{56}$Fe and their fragments with charge greater than $Z=19$.}
\end{center}
\end{figure}

Following the same procedure, the dependence of etched track area on the charge of beam and fragments for fragmentation of $^{56}$Fe on other targets are given in follows:

\begin{eqnarray}
Area=-242.54+280.01Z & \mbox{$\chi^{2}_{min}=0.06$, 496 A MeV $^{56}$Fe+CH$_{2}$}\\
Area=83.71+267.29Z   & \mbox{$\chi^{2}_{min}=0.04$, 496 A MeV $^{56}$Fe+C}\\
Area=53.10+266.83Z   & \mbox{$\chi^{2}_{min}=0.03$, 468 A MeV $^{56}$Fe+C}\\
Area=-627.21+294.26Z & \mbox{$\chi^{2}_{min}=0.23$, 496 A MeV $^{56}$Fe+Al}\\
Area=-393.82+278.43Z  & \mbox{$\chi^{2}_{min}=0.01$, 477 A MeV $^{56}$Fe+Al}
\end{eqnarray}

Based on the dependence of etched track area on the charge of beam and fragments for fragmentation of $^{56}$Fe on different targets at different beam energies, the charge of projectile fragments are identified. Fig.5 shows the projectile fragment etched track area distribution for fragmentation of $^{56}$Fe on different targets at different beam energies. The distribution with the highest peak is fragment with charge $Z=25$. Following the highest peak, the other peak represent the distribution of projectile fragments with charge of $Z=24, 23, 22$ and so on.

\begin{figure}[htbp]
\begin{center}
\includegraphics[width=0.60\linewidth]{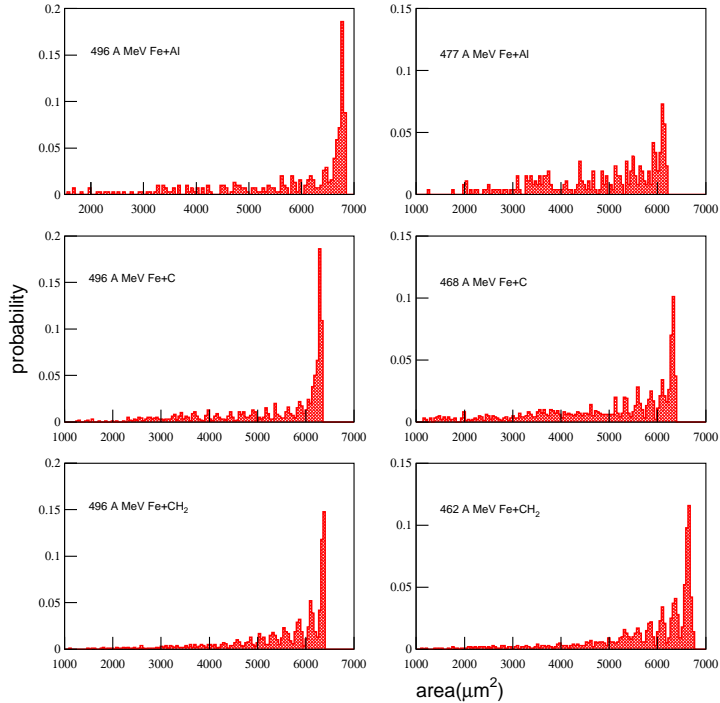}
\caption{(Color online) The etched track area distribution of projectile fragments for fragmentation of $^{56}$Fe on different targets at different beam energies.}
\end{center}
\end{figure}

\section*{3. RESULTS AND DISCUSSION}

The emission angles of PFs after each targets related to the direction of $^{56}$Fe ions before the target are determined using the coordinates of  $^{56}$Fe ions, PFs and the thickness of detector. Track dimensions and positions within one microscope frame can be measured with an accuracy of $\sim0.1\mu$m. However, overall positional accuracy is dominated by the moving stage of the microscope. The positional uncertainty $\sigma_{p}$ in the $x-y$ plane of the stack coordinate system is about 3 $\mu$m. The positional uncertainty $\sigma_{z}$ in the z-axis comes from stack composition and detector thickness measurement and is estimated at $\sim8\mu$m. Using the quadruplet fitting method the corresponding angular uncertainty is
\begin{equation}
\sigma(\theta)=\frac{\sqrt{\sigma_{z}^{2}\sin^{2}\theta+2\sigma_{p}^{2}\cos^{2}\theta}}{2h},
\end{equation}
where $\theta$ represents the polar angle of the fitted line. With a detector thickness of $h=780\mu$m we thus obtain angular uncertainties $\sigma(\theta)\approx0.16^{\circ}$ for the value of $\theta$ up to 8$^{\circ}$.

\begin{table}
\begin{center}
Table 1.The mean emission angles(unit: degree) of PFs for $^{56}$Fe fragmentation on different targets at different energies(in small brackets, unit: A MeV).\\
\begin{small}
\begin{tabular}{ccccccc}\hline
Charge &\multicolumn{2}{c}{Al-target} & \multicolumn{2}{c}{C-target} & \multicolumn{2}{c}{CH$_{2}$-target} \\
of PFs & $<\theta>$ (496) & $<\theta>$ (477) & $<\theta>$ (496) & $<\theta>$ (468) & $<\theta>$ (496) &$<\theta>$ (462) \\\hline
$Z=25$ & $0.535\pm0.105$ & $0.554\pm0.095$ & $0.145\pm0.035$ & $0.271\pm0.045$ & $0.136\pm0.005$ & $0.240\pm0.018$ \\
$Z=24$ & $0.209\pm0.039$ & $0.427\pm0.112$ & $0.150\pm0.014$ & $0.168\pm0.017$ & $0.168\pm0.008$ & $0.186\pm0.011$ \\
$Z=23$ & $0.375\pm0.080$ & $0.232\pm0.050$ & $0.250\pm0.026$ & $0.215\pm0.023$ & $0.234\pm0.013$ & $0.276\pm0.016$ \\
$Z=22$ & $0.418\pm0.102$ & $0.307\pm0.041$ & $0.268\pm0.033$ & $0.250\pm0.031$ & $0.269\pm0.017$ & $0.248\pm0.021$ \\
$Z=21$ & $0.383\pm0.076$ & $0.321\pm0.040$ & $0.314\pm0.034$ & $0.307\pm0.038$ & $0.244\pm0.020$ & $0.342\pm0.023$ \\
$Z=20$ & $0.387\pm0.086$ & $0.559\pm0.154$ & $0.338\pm0.040$ & $0.393\pm0.043$ & $0.359\pm0.029$ & $0.363\pm0.028$ \\
$Z=19$ & $0.348\pm0.163$ & $0.378\pm0.095$ & $0.487\pm0.057$ & $0.425\pm0.066$ & $0.413\pm0.036$ & $0.419\pm0.042$ \\
$Z=18$ & $0.459\pm0.085$ & $0.609\pm0.093$ & $0.433\pm0.052$ & $0.576\pm0.069$ & $0.448\pm0.035$ & $0.468\pm0.051$ \\
$Z=17$ & $0.461\pm0.131$ & $1.214\pm0.417$ & $0.454\pm0.072$ & $0.520\pm0.058$ & $0.577\pm0.070$ & $0.503\pm0.062$ \\
$Z=16$ & $1.009\pm0.211$ & $0.591\pm0.185$ & $0.475\pm0.081$ & $0.806\pm0.108$ & $0.491\pm0.052$ & $0.587\pm0.062$ \\
$Z=15$ & $0.709\pm0.213$ & $0.970\pm0.169$ & $0.731\pm0.093$ & $0.668\pm0.077$ & $0.607\pm0.081$ & $0.588\pm0.076$ \\
$Z=14$ & $0.721\pm0.232$ & $0.587\pm0.115$ & $0.606\pm0.077$ & $0.849\pm0.083$ & $0.477\pm0.063$ & $0.841\pm0.147$ \\
$Z=13$ & $0.550\pm0.097$ & $0.784\pm0.115$ & $0.850\pm0.095$ & $0.711\pm0.092$ & $0.702\pm0.082$ & $0.755\pm0.112$ \\
$Z=12$ & $0.713\pm0.197$ & $0.807\pm0.018$ & $0.779\pm0.134$ & $0.807\pm0.103$ & $0.850\pm0.172$ & $1.009\pm0.168$ \\
$Z=11$ & $1.820\pm0.805$ & $1.214\pm0.417$ & $1.102\pm0.161$ & $1.081\pm0.238$ & $0.788\pm0.113$ & $0.783\pm0.199$ \\
$Z=10$ & $1.363\pm0.825$ & $2.199\pm2.199$ & $0.935\pm0.110$ & $1.523\pm0.336$ & $0.907\pm0.276$ & $1.245\pm0.195$ \\
$Z=9$ & $0.943\pm0.063$  &                 & $0.807\pm0.107$ & $0.898\pm0.172$ & $1.301\pm0.291$ & $1.165\pm0.273$ \\
$Z=8$ & $1.108\pm1.108$  &                 & $0.339\pm0.029$ & $0.653\pm0.260$ & $2.705\pm1.559$ & $1.029\pm0.264$ \\
$Z=7$ &                  &                 & $0.910\pm0.910$ & $1.266\pm0.344$ & $0.778\pm0.778$ & $0.995\pm0.607$ \\
$Z=6$ &                  &                 & $0.179\pm0.179$ & $1.213\pm0.398$ & $0.677\pm0.677$ & $0.697\pm0.067$ \\\hline
\end{tabular}
\end{small}
\end{center}
\end{table}

Fig.6 shows the angular distribution of PFs from the fragmentation of $^{56}$Fe on Al target at 496 A MeV (a) and 477 A MeV (b). Fig.7 shows the angular distribution of PFs from the fragmentation of $^{56}$Fe on C target at 496 A MeV (a) and 468 A MeV (b). Fig.8 shows the angular distribution of PFs from the fragmentation of $^{56}$Fe on CH$_{2}$ target at 496 A MeV (a) and 462 A MeV (b). From these figures we can see that most PFs have a emission angle less than 1.0 degree, a little of them is great than 1.0 degree. With the decrease of the charge of PF, the angular distribution are widened. The angular distribution do not obviously depend on the target mass for the same PF. The mean emission angles of PFs for $^{56}$Fe fragmentation on different targets at different energies are presented in Table 1. The mean emission angle distribution of PFs produced in fragmentation of $^{56}$Fe on different targets at different energies is shown in Fig. 9. It is found that the mean emission angle increases with the decrease of charge of PF, and no obvious beam energy and target size dependence are found in our studied beam energy region.

\begin{figure}[htbp]
\begin{center}
\includegraphics[width=0.49\linewidth]{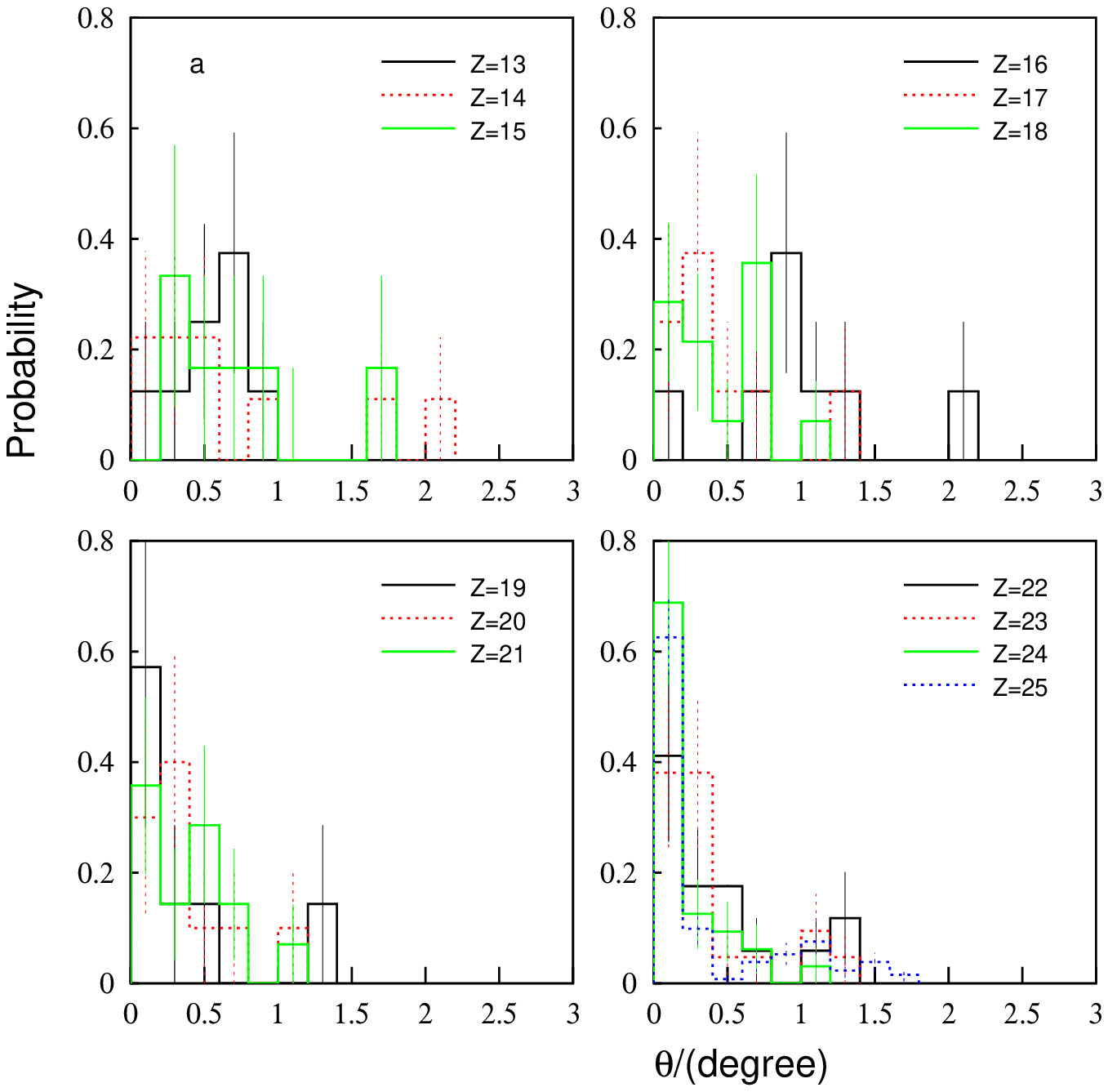}
\includegraphics[width=0.49\linewidth]{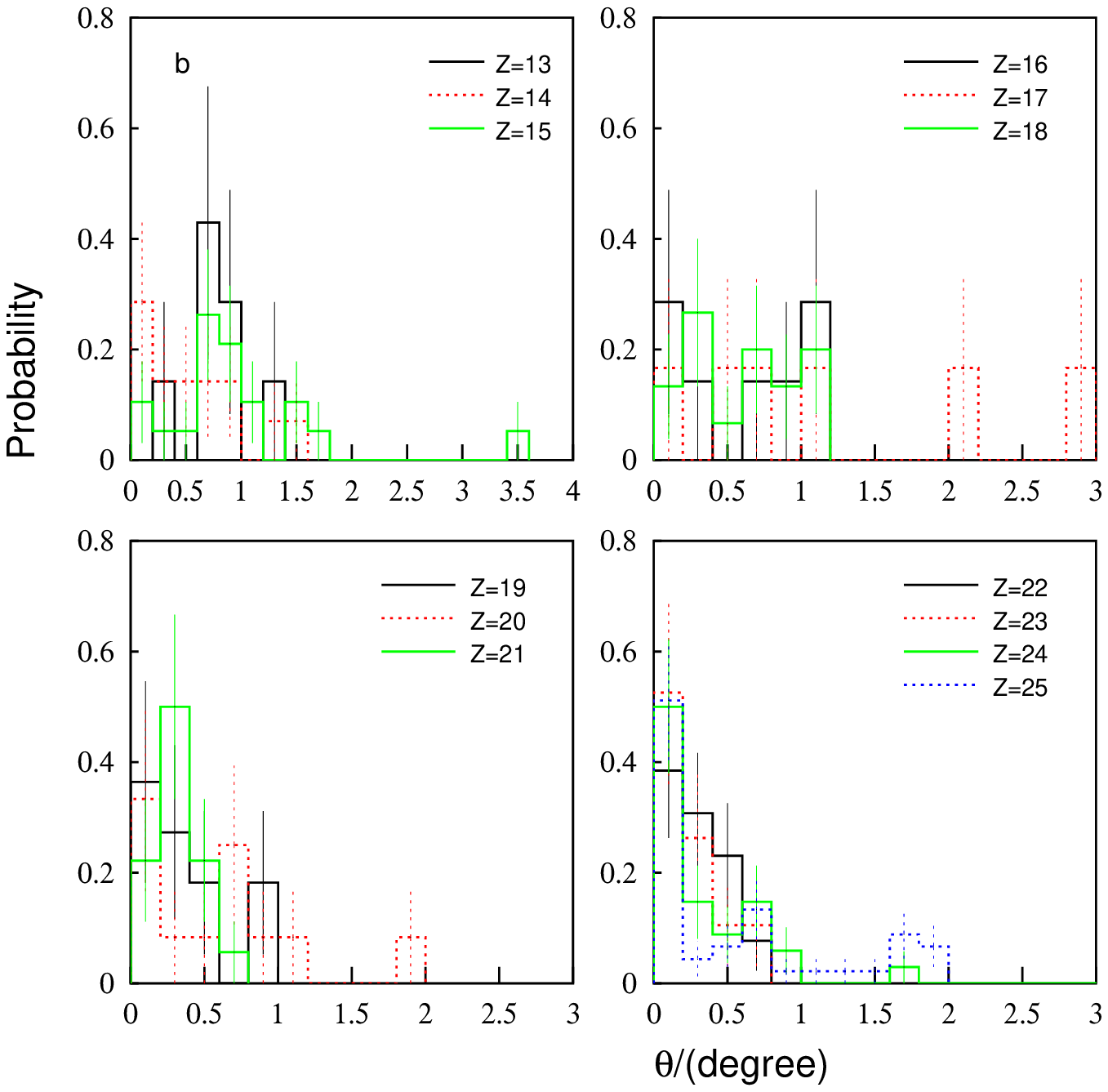}
\caption{(Color online) The angular distribution of PFs from the fragmentation of $^{56}$Fe on Al target at 496 A MeV (a) and 477 A MeV (b).}
\end{center}
\end{figure}

\begin{figure}[htbp]
\begin{center}
\includegraphics[width=0.49\linewidth]{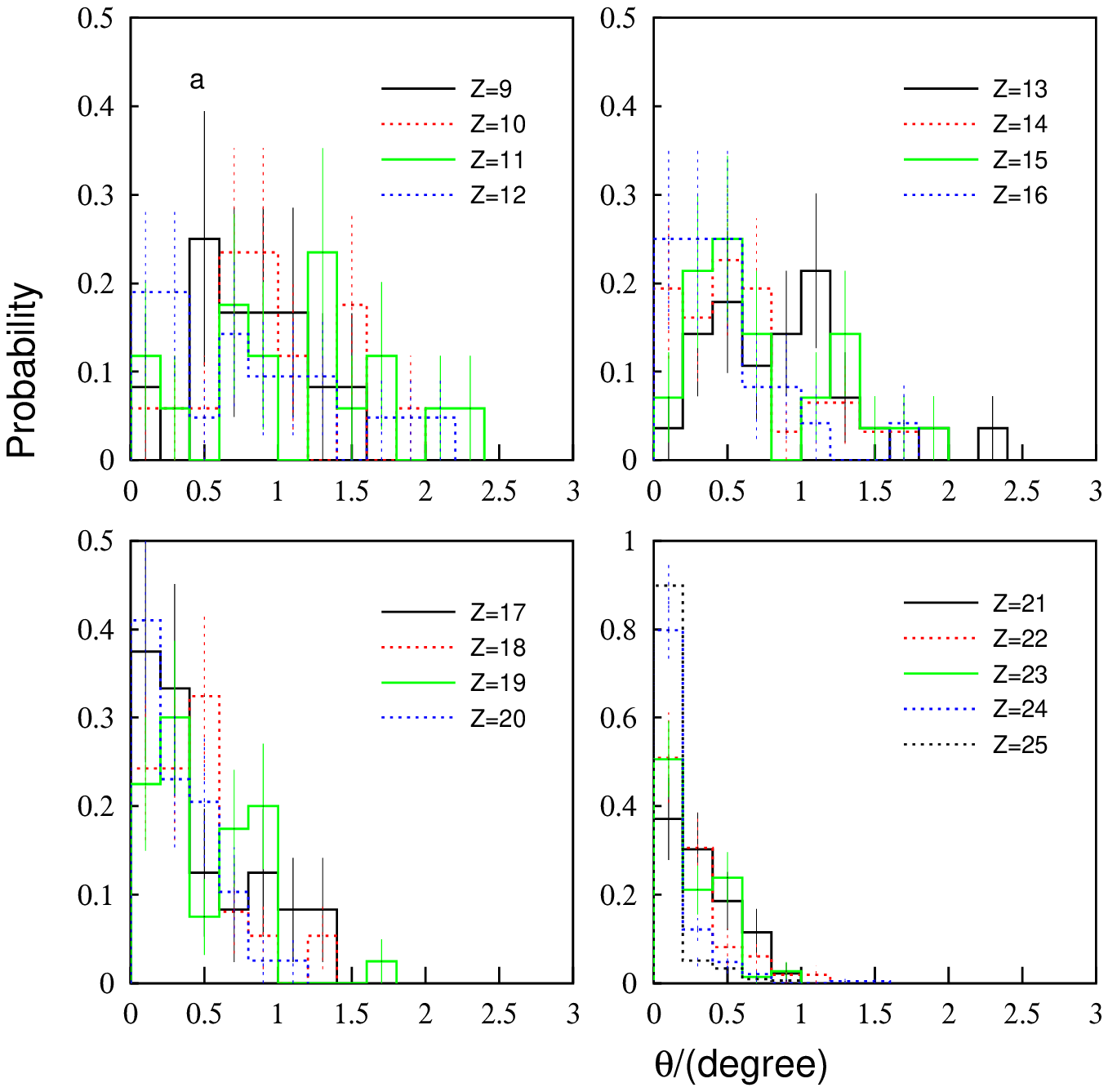}
\includegraphics[width=0.49\linewidth]{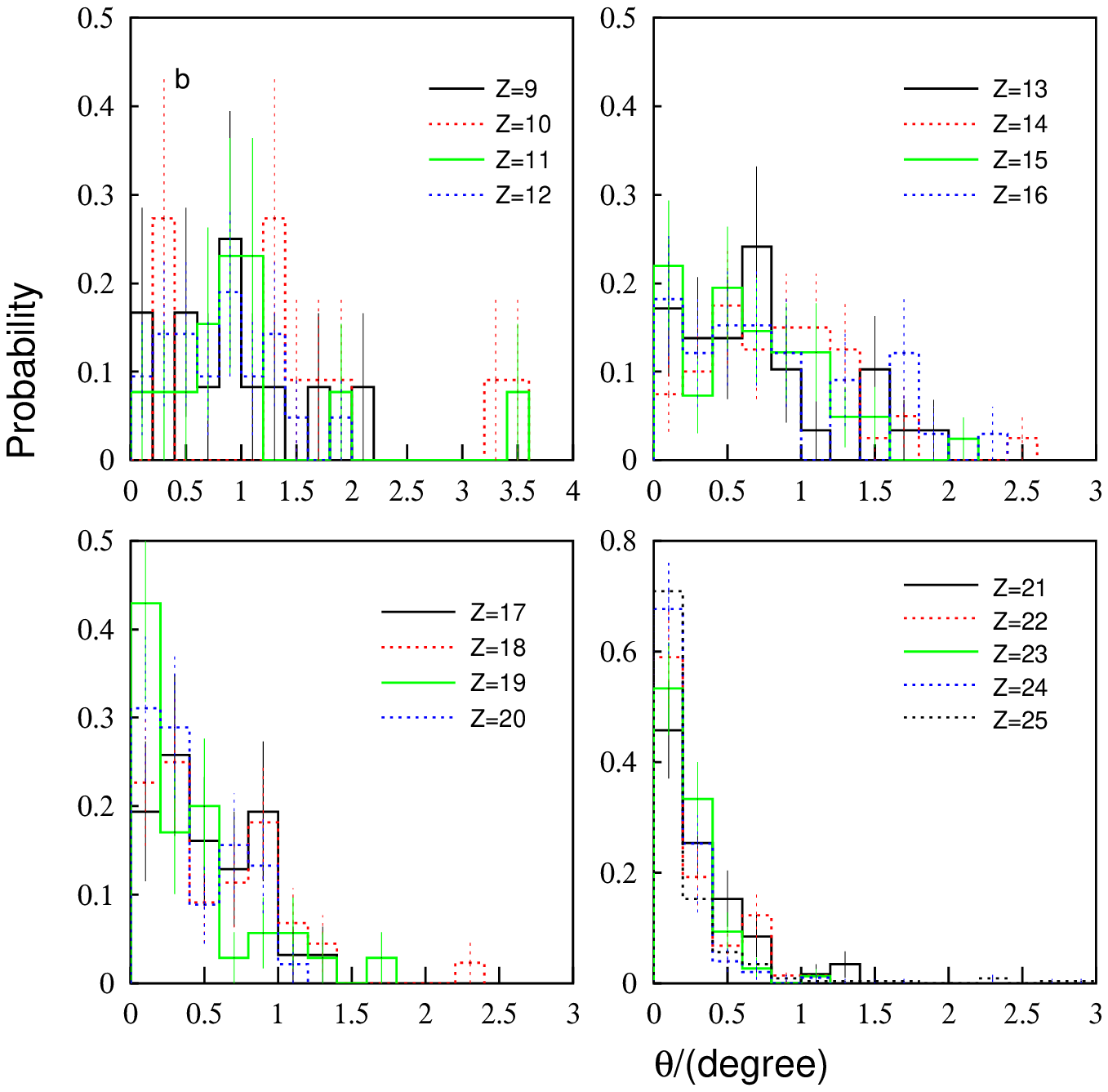}
\caption{(Color online) The angular distribution of PFs from the fragmentation of $^{56}$Fe on C target at 496 A MeV (a) and 468 A MeV (b).}
\end{center}
\end{figure}

\begin{figure}[htbp]
\begin{center}
\includegraphics[width=0.49\linewidth]{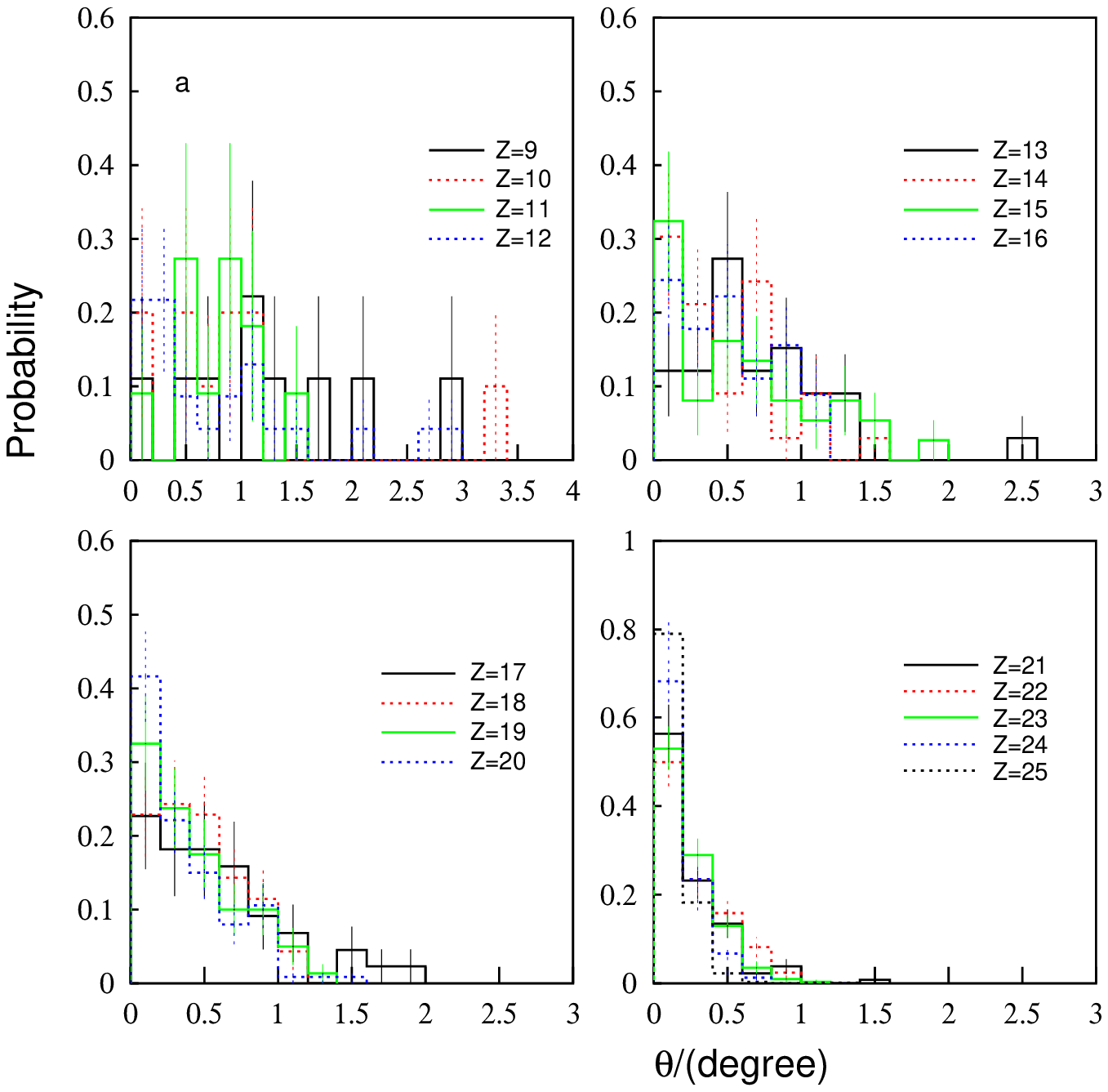}
\includegraphics[width=0.49\linewidth]{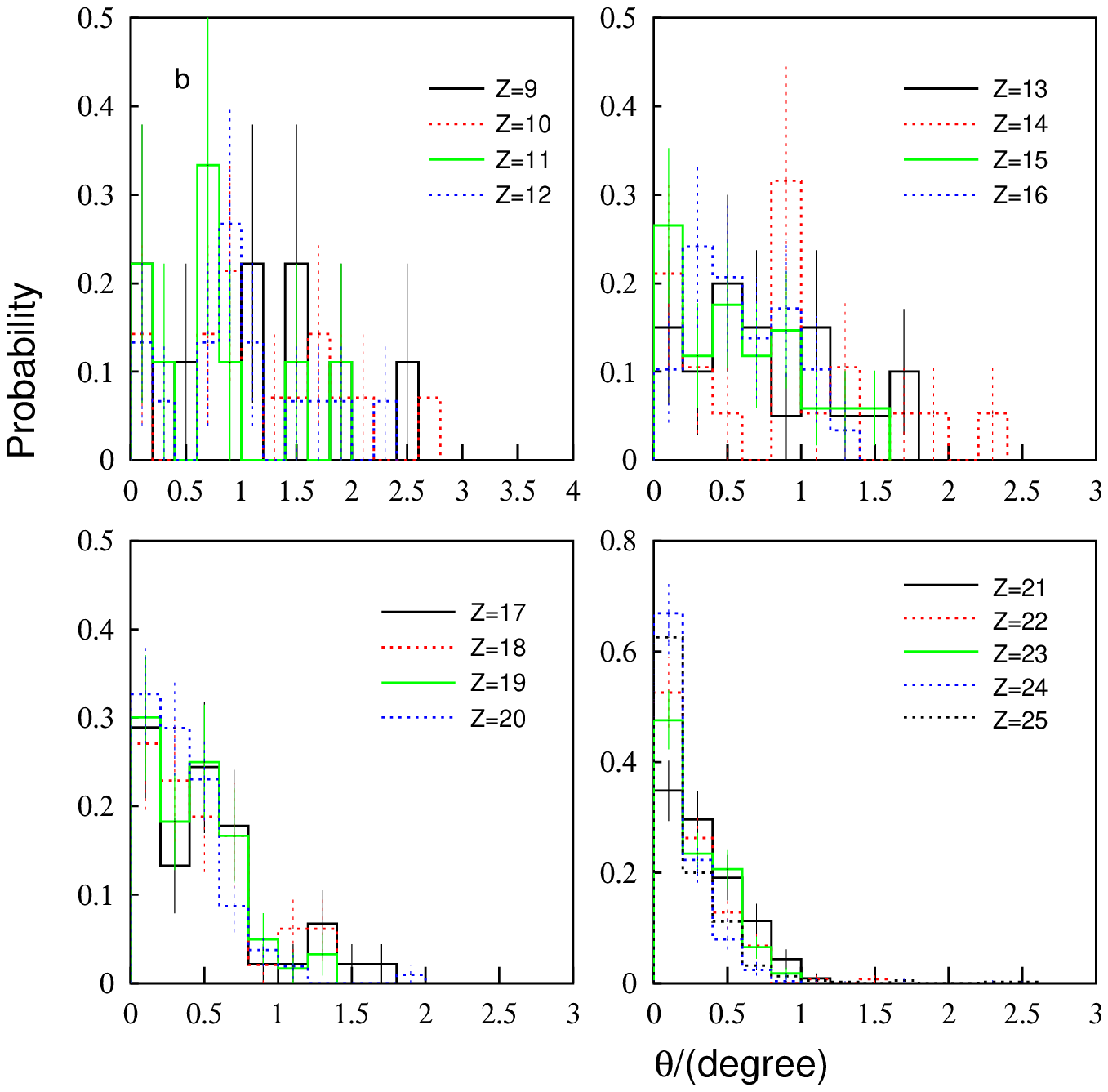}
\caption{(Color online) The angular distribution of PFs from the fragmentation of $^{56}$Fe on CH$_{2}$ target at 496 A MeV (a) and 462 A MeV (b).}
\end{center}
\end{figure}

\begin{figure}[htbp]
\begin{center}
\includegraphics[width=0.60\linewidth]{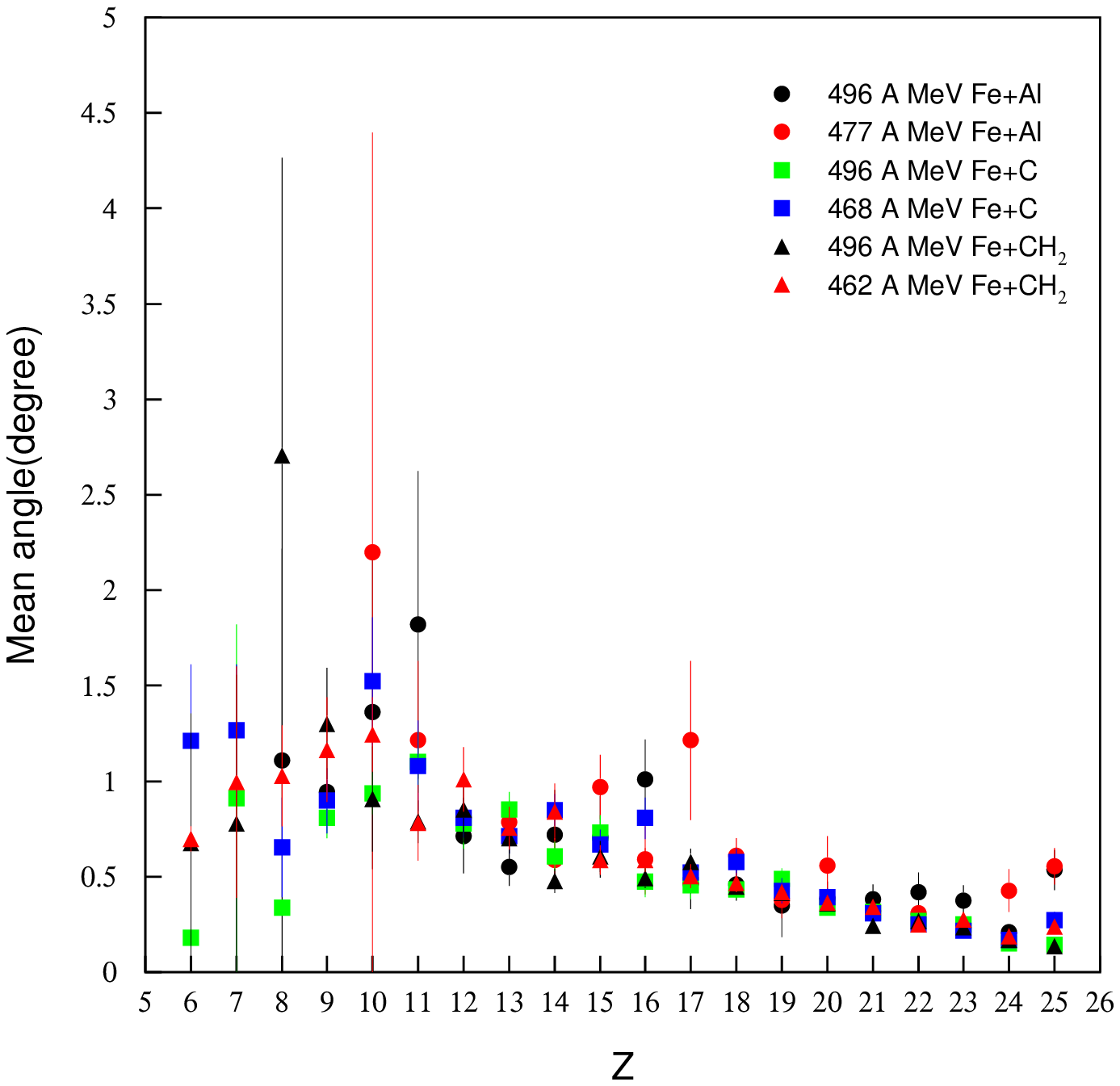}
\caption{(Color online) The dependence of mean emission angle on the charge of PFs for the fragmentation of $^{56}$Fe on different targets at different beam energy.}
\end{center}
\end{figure}

The transverse momentum per nucleon ($p_{t}$) of PF is calculated on the basis of its emission angle, $p_{t}=p\sin{\theta}$, where $p$ is the momentum per nucleon of beam which can be calculated from the beam energy per nucleon ($E$), $p=\sqrt{E^{2}+2m_{0}E}$, $m_{0}$ is the nucleon rest mass.

The spectator is that part of the system which has not collided with the other nucleus, but which is nevertheless excited due to the shearing-off of part of the nucleus and due to absorption of participant particles. The projectile spectator is identified as those particles which have approximately beam energy per nucleon but have a small reflecting angle related to the beam direction. It was seen in Ref.[18] that it represents a well equilibrated piece of nuclear matter at finite temperature. When the spectator is fully developed the properties are rather independent of incident energy which supports the freeze-out picture[19], and the density and temperature remain rather constant for several tens of fm/c, making it an ideal system in order to study the thermodynamical evolution of low-density, finite temperature nuclear matter. According to the participant-spectator concept and the fireball model[20], if we assume that the emission of PFs is Maxwell-Boltzmann distribution in projectile rest frame with a certain temperature {\em T}, then the integral frequency distribution of the squared transverse momentum per nucleon is
\begin{equation}
\ln{F}(>{p^{2}_{t}})=-Ap^{2}_{t}/2m_{p}T
\end{equation}
where A is the mass number of PF, $m_{p}$ is the mass of proton. The linearity of such a plot would be strong evidence for a single temperature of emission source. Fig.9 shows the cumulative plots of {\em F} as a function of $p^{2}_{t}$ for PFs from the fragmentation of $^{56}$Fe on Al target at 496 A MeV (a) and 477 A MeV (b). Fig.10 shows the cumulative plots of {\em F} as a function of $p^{2}_{t}$ for PFs from the fragmentation of $^{56}$Fe on C target at 496 A MeV (a) and 468 A MeV (b). Fig.11 shows the cumulative plots of {\em F} as a function of $p^{2}_{t}$ for PFs from the fragmentation of $^{56}$Fe and on CH$_{2}$ target at 496 A MeV (a) and 462 A MeV. All of the plots can be fitted by a single Rayleigh distribution of the form
\begin{equation}
\ln{F}(>{p^{2}_{t}})=C\exp{-p^{2}_{t}/2\sigma^{2}}
\end{equation}
where $\sigma=\sqrt{2/\pi}<p_{t}>$, which is related to the temperature of PF emission source. Using the fitting parameter the temperature of PF emission source is calculated which is presented in Table 2. It is shown that the temperature of PF emission source does not obviously depend on target size. Average speaking, the temperature of heavier PFs emission source is less than that of lighter PFs emission source, but the difference is not so obvious. The temperature of PF emission source is about 1-8 MeV for the PFs with charge in the range from 9 to 25, which is in good agreement with the finding of Ref.[11-13,16] based on isotope thermometers but less than the result of Ref.[8] based on the PF kinetic energy spectrum.

\begin{figure}[htbp]
\begin{center}
\includegraphics[width=0.49\linewidth]{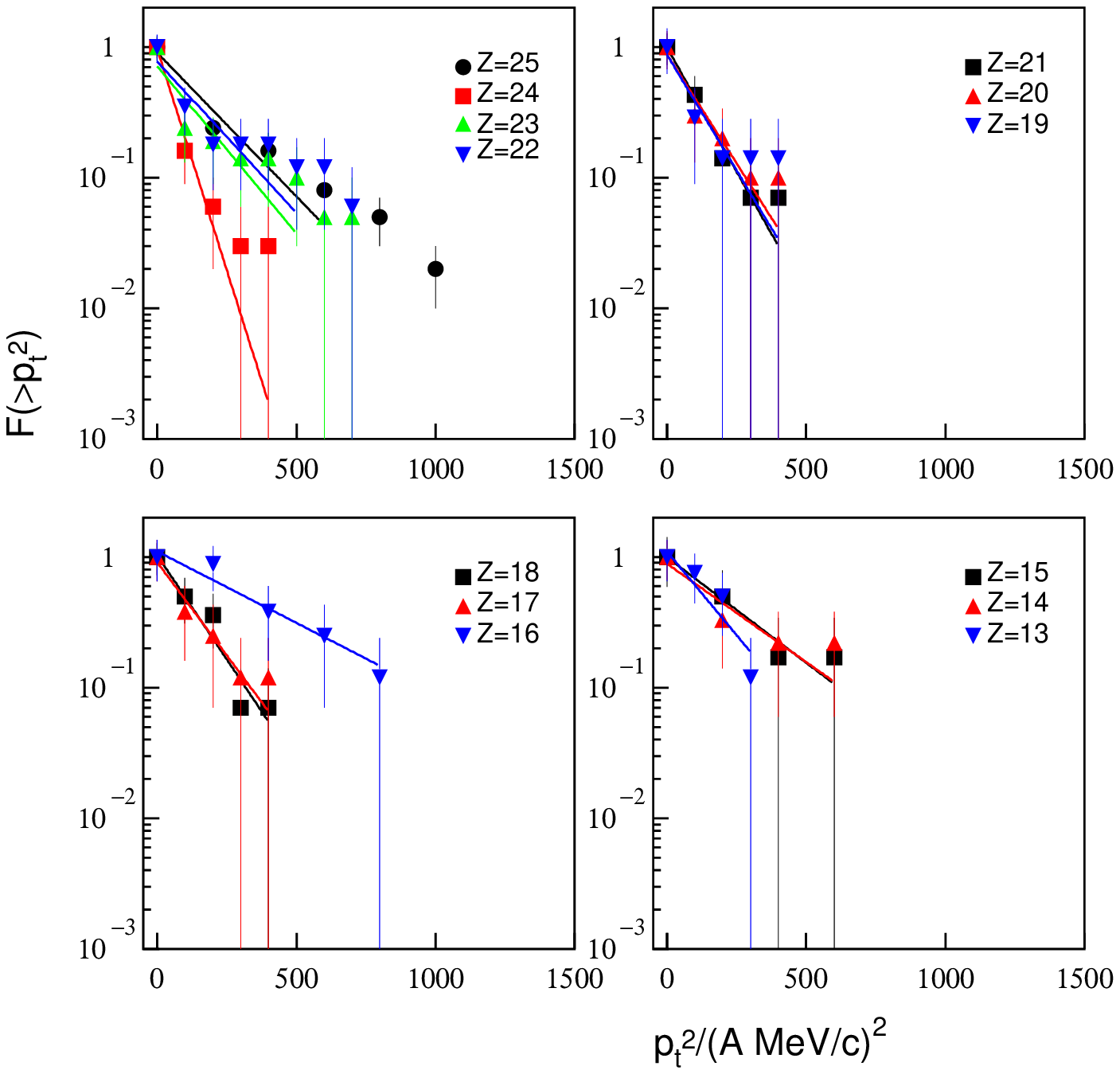}
\includegraphics[width=0.49\linewidth]{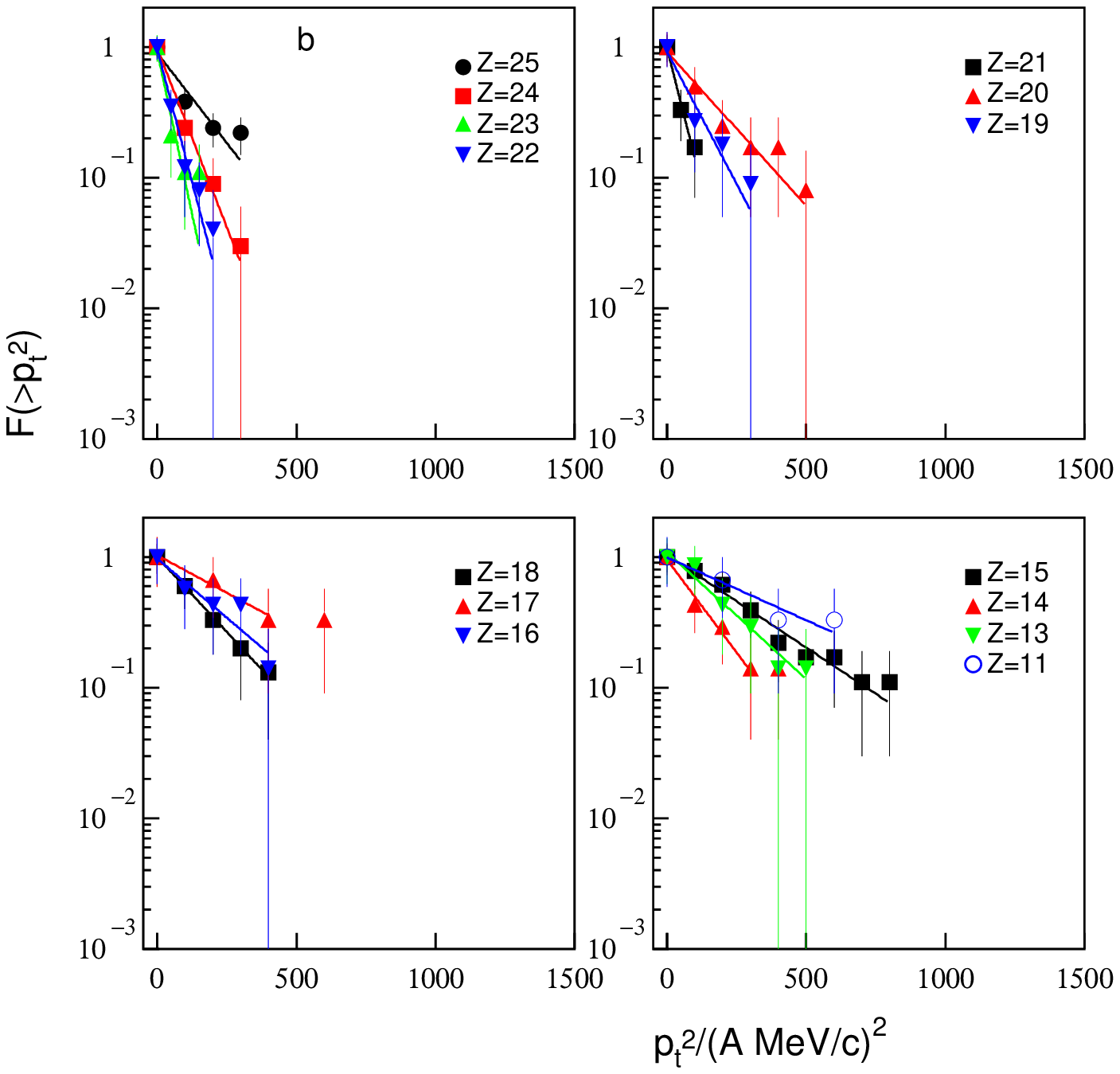}
\caption{(Color online) The cumulative $p^{2}_{t}$ distribution of PFs from the fragmentation of $^{56}$Fe on aluminum target at 496 A MeV (a) and 477 A MeV (b).}
\end{center}
\end{figure}

\begin{figure}[htbp]
\begin{center}
\includegraphics[width=0.49\linewidth]{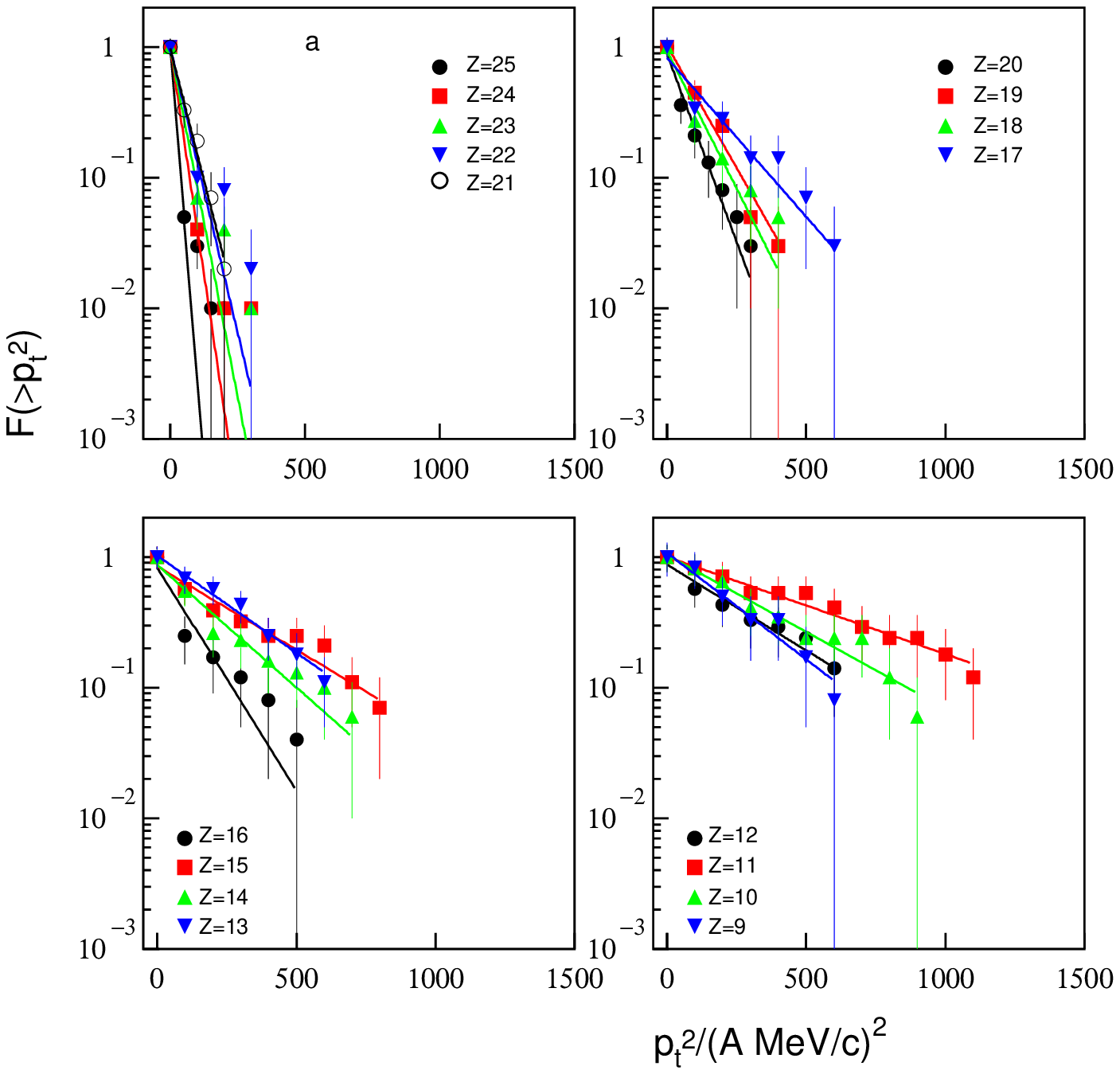}
\includegraphics[width=0.49\linewidth]{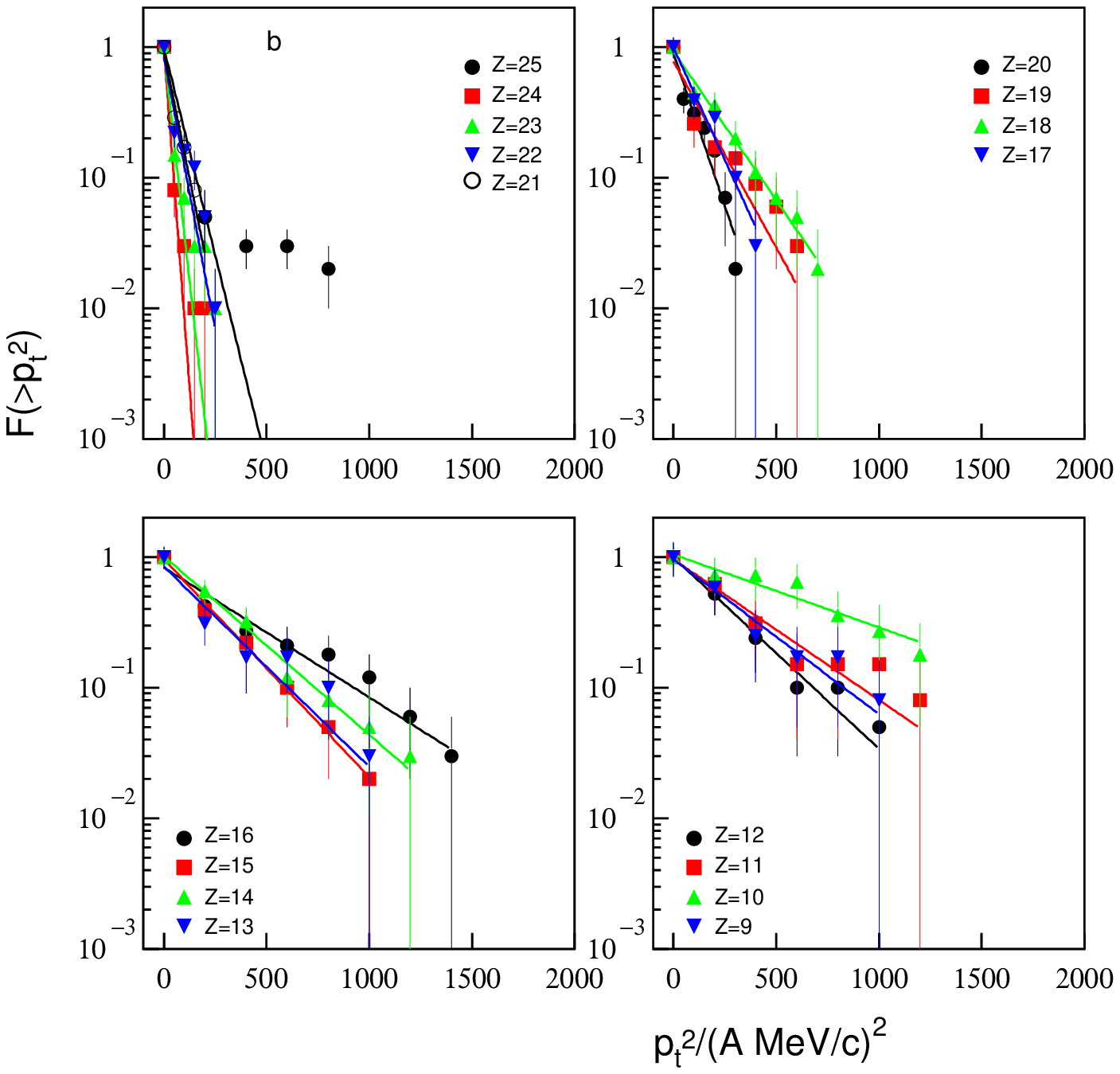}
\caption{(Color online) The cumulative $p^{2}_{t}$ distribution of PFs from the fragmentation of $^{56}$Fe on carbon target at 496 A MeV (a) and 468 A MeV (b).}
\end{center}
\end{figure}

\begin{figure}[htbp]
\begin{center}
\includegraphics[width=0.49\linewidth]{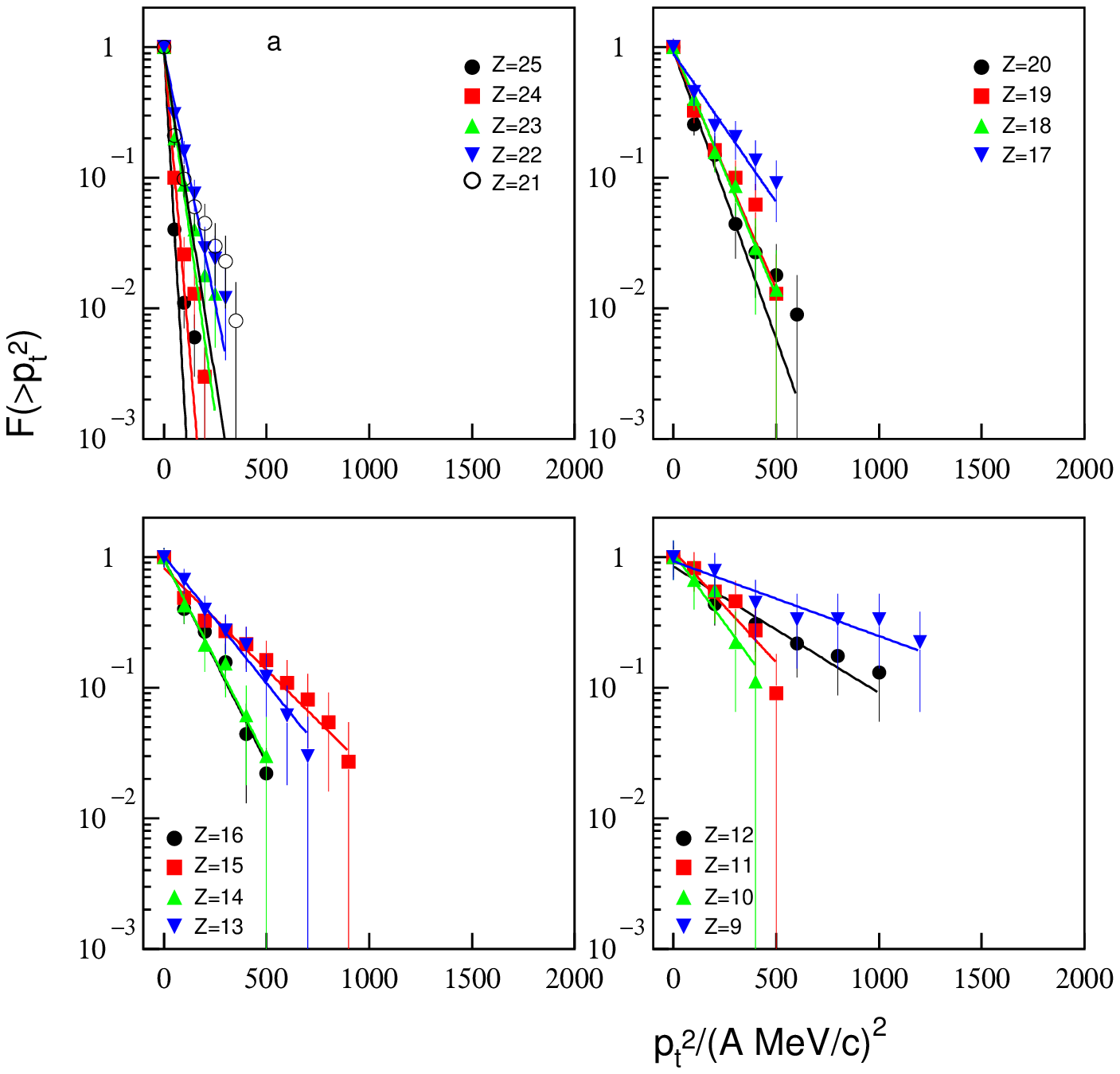}
\includegraphics[width=0.49\linewidth]{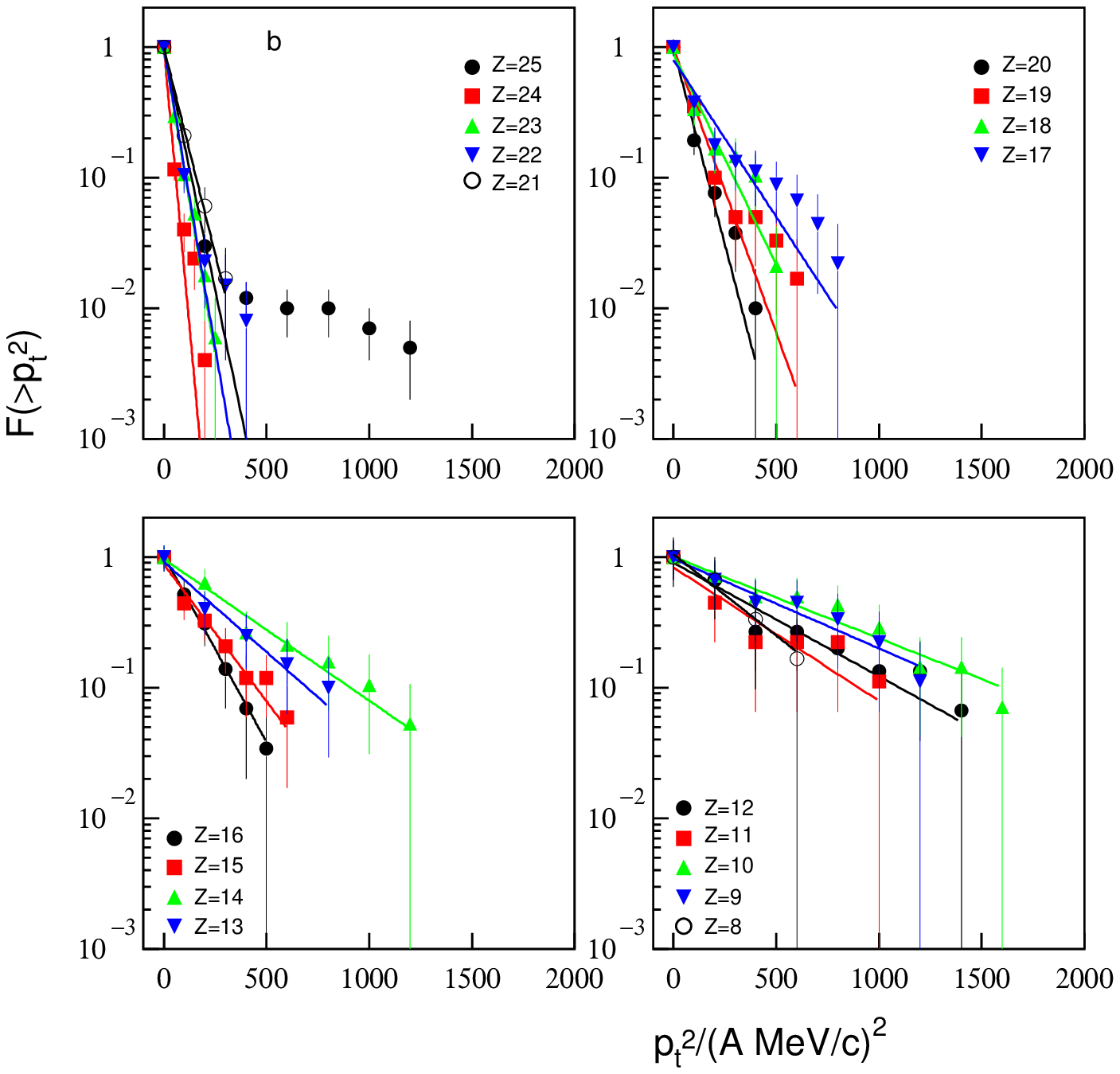}
\caption{(Color online) The cumulative $p^{2}_{t}$ distribution of PFs from the fragmentation of $^{56}$Fe on polyethylene target at 496 A MeV (a) and 462 A MeV (b).}
\end{center}
\end{figure}

The PFs are produced from the peripheral heavy ion collisions, which provides an ideal scenario for studying multifragment decay of hot and dilute nuclei. According to the participant-spectator concept[20], it is assumed that when the interaction of projectile and target nuclei takes place, the projectile and target sweep out cylindrical cuts through each other. During the separation of the spectators from the participants, there is some intercommunication, which results in the excitation of the spectators. This excitation strongly depends on the contacted area of the colliding system. The heavier fragment corresponds to the large impact parameter and the small contacted areas, the lighter fragment corresponds to the smaller impact parameter and the larger contacted areas. So the excitation energy of the heavier fragments less than that of the lighter fragments, results in the temperature of an emission source of a heavier fragment being less than that of the light fragment.

\begin{table}
\begin{center}
Table 2.The temperature (unit: MeV) of PF emission source for $^{56}$Fe fragmentation on different targets at different energies(in small brackets, unit: A MeV).\\
\begin{small}
\begin{tabular}{ccccccc}\hline
Charge &\multicolumn{2}{c}{Al-target} & \multicolumn{2}{c}{C-target} & \multicolumn{2}{c}{CH$_{2}$-target} \\
of PFs & T (496) & T (477) & T (496) & T (468) & T (496)& T (462) \\\hline
$Z=25$ & $5.68\pm0.74$ & $4.63\pm1.07$ & $0.50\pm0.07$ & $2.00\pm0.14$ & $0.47\pm0.03$ & $1.71\pm0.11$ \\
$Z=24$ & $1.05\pm0.24$ & $2.20\pm0.43$ & $0.87\pm0.07$ & $0.58\pm0.09$ & $0.65\pm0.05$ & $0.70\pm0.06$ \\
$Z=23$ & $3.46\pm2.61$ & $1.19\pm0.39$ & $1.10\pm0.18$ & $0.83\pm0.13$ & $1.06\pm0.09$ & $1.26\pm0.10$ \\
$Z=22$ & $5.23\pm1.99$ & $1.37\pm0.29$ & $1.28\pm0.29$ & $1.33\pm0.26$ & $1.44\pm0.12$ & $1.21\pm0.14$ \\
$Z=21$ & $1.81\pm0.48$ & $1.24\pm0.42$ & $1.29\pm0.19$ & $1.34\pm0.22$ & $1.04\pm0.14$ & $1.65\pm0.17$ \\
$Z=20$ & $1.84\pm0.76$ & $3.94\pm1.29$ & $1.61\pm0.28$ & $1.98\pm0.26$ & $2.10\pm0.23$ & $1.55\pm0.19$ \\
$Z=19$ & $2.22\pm1.18$ & $2.23\pm0.94$ & $2.41\pm0.34$ & $3.17\pm0.82$ & $2.48\pm0.30$ & $2.07\pm0.29$ \\
$Z=18$ & $2.81\pm0.69$ & $4.04\pm1.13$ & $2.20\pm0.47$ & $4.07\pm0.50$ & $2.43\pm0.26$ & $2.87\pm0.49$ \\
$Z=17$ & $1.77\pm0.87$ & $7.06\pm4.91$ & $3.33\pm0.69$ & $2.36\pm0.40$ & $3.56\pm0.63$ & $3.38\pm0.71$ \\
$Z=16$ & $6.17\pm1.63$ & $4.09\pm1.84$ & $2.18\pm0.70$ & $7.45\pm1.27$ & $2.37\pm0.21$ & $2.61\pm0.41$ \\
$Z=15$ & $4.22\pm2.24$ & $5.07\pm0.98$ & $5.62\pm1.02$ & $4.31\pm0.59$ & $4.60\pm0.66$ & $3.42\pm0.61$ \\
$Z=14$ & $4.31\pm2.41$ & $2.28\pm0.78$ & $3.46\pm0.67$ & $4.75\pm0.63$ & $2.13\pm0.36$ & $5.98\pm1.31$ \\
$Z=13$ & $1.66\pm0.50$ & $3.22\pm1.09$ & $4.21\pm0.70$ & $4.08\pm1.02$ & $3.20\pm0.41$ & $4.53\pm1.23$ \\
$Z=12$ &               &               & $4.23\pm1.10$ & $3.79\pm0.80$ & $5.74\pm1.56$ & $6.40\pm1.71$ \\
$Z=11$ &               & $5.54\pm3.55$ & $7.17\pm1.33$ & $4.95\pm1.44$ & $3.12\pm0.80$ & $5.21\pm2.48$ \\
$Z=10$ &               &               & $3.93\pm0.70$ & $8.30\pm2.52$ & $2.15\pm0.68$ & $7.43\pm1.54$ \\
$Z=9$ &                &               & $2.70\pm0.65$ & $3.71\pm1.14$ & $7.69\pm3.32$ & $6.43\pm2.13$ \\
$Z=8$ &                &               &               &               &         & $2.98\pm1.40$ \\\hline
\end{tabular}
\end{small}
\end{center}
\end{table}

\section*{4. CONCLUSIONS}
The emission angle distributions of PFs and the temperature of PF emission source for fragmentation of $^{56}$Fe on C, Al and CH$_{2}$ targets at different energies are studied. It is found that the averaged emission angle increase with the decrease of the charge of PF for the same target, and no obvious dependence of angular distribution on the mass of target nucleus is found for the same PF. The cumulated squared transverse momentum distribution of PF can be well represented by a single Rayleigh distribution, the temperature parameter of PFs emission source is obtained, which is about $1.0\sim8.0$ MeV and do not depend on the mass of target for PF with charge of $9\leq{Z}\leq25$. Average speaking, the temperature of heavier PFs emission source is less than that of lighter PFs emission source, but the difference is not so obvious.

\section*{Acknowledgment}
This work has been supported by the Chinese National Science Foundation under Grant Nos: 11075100 and 11565001, the Natural Foundation of Shanxi Province under Grant 2011011001-2, the Shanxi Provincial Foundation for Returned Overseas Chinese Scholars, China (Grant No. 2011-058). We are grateful to staff of the HIMIC for helping to expose the stacks.

\end{document}